\documentclass[fleqn,twoside]{article} 
\usepackage{epsf,multicol,ifthen}
\usepackage{ujp}
\usepackage[dvips]{graphicx}

\usepackage[T2A]{fontenc}
\usepackage[cp1251]{inputenc}
\usepackage[english,russian]{babel}
\usepackage{amsmath, amssymb}
\usepackage{cite}

\makeatother

\newcommand\I{\texttt{i}}
\newcommand\D{\mathrm{d}}
\newcommand\E{\mathrm{e}}
\newcommand\refe[1]{(\ref{#1})}
\nazva{TWO-PARAMETER METHOD\\ FOR DESCRIBING  THE NONLINEAR\\ EVOLUTION  OF NARROW-BAND WAVE TRAINS}%

\udk{530.182}

\nazvacol{TWO-PARAMETER METHOD FOR DESCRIBING  THE NONLINEAR EVOLUTION}%

\avtor{\fbox{V.P. LUKOMSKY}, I.S. GANDZHA}%
\avtorcol{V.P. LUKOMSKY, I.S. GANDZHA}%

\inst{Institute of Physics of the NAS of Ukraine}%
\adr{(46, Nauky Prosp., Kyiv 03028, Ukraine; e-mail: gandzha@iop.kiev.ua)}

\begin{document}

\setcounter{page}{207}%
\maketitl                 
\begin{multicols}{2}
\anot{We consider the evolution of
narrow-band wave trains of finite amplitude in a nonlinear
dispersive system which is described by the Klein--Gordon equation
with arbitrary polynomial nonlinearity. We use a new perturbative
technique which allows the original wave equation to be reduced to a
model equation for the wave train envelope (high-order nonlinear
Schr\"{o}dinger equation). The time derivative is expanded into an
asymptotic series in two independent parameters which
characterize the smallness of amplitudes ($\varepsilon$) and the
slowness of their spatial variations ($\mu$). In contrast to other
perturbative methods in which these two parameters are taken equal
(e.g., the multiple scale method), the two-parameter method produces
no secular terms. The results of this study can be applied to
investigating the propagation of ultrashort (femtosecond) pulses in
optical fibers,
to studying the wave events on a fluid surface, and to describing the Langmuir waves in hot plasmas.}%

\section{Introduction}
One of the remarkable properties of nonlinear dispersive systems is
the possibility of the existence of steady progressive waves with
finite amplitude due to the balance between dispersion and
nonlinearity
effects. 
In the absence of dissipation, the dispersion relation (dependence
of wave frequency on wavelength) can be written as
\begin{equation}\label{eq:dispersion}
\omega=\omega(k,\;|A|^2),
\end{equation}
where $k \equiv 2\pi/\lambda$ is the wave number, $\omega$ is the
wave frequency, and $\lambda$ is the wavelength. In systems with no
dispersion, the dependence on $k$ is linear, i.e., $\omega = ck$.
The constant $c$ is called the wave phase speed. In the case of
nonzero dispersion, the phase speed depends on wave number
($\partial^2\omega/\partial k^2\neq0$), so that wave trains spread
out in space. In nonlinear systems, this dispersive spreading can be
compensated  by nonlinear effects, which manifest themselves in the
dependence of wave frequency on wave amplitude $A$ in dispersion
relation \refe{eq:dispersion}.

For narrow-band wave trains, when $\Delta k\ll k$, the dispersion
relation can be expanded into a Taylor series about the carrier
wave number $k_0$ and frequency $\omega_0$:
\[
\omega-\omega_0\equiv\Delta\omega=\left(\frac{\partial\omega}{\partial
k}\right)_{k=k_0}\kern-1em\Delta
k+\frac{1}{2}\left(\frac{\partial^2\omega}{\partial
k^2}\right)_{k=k_0}\kern-1em\Delta k^2+\]
\begin{equation}
+\left(\frac{\partial\omega}{\partial
|A|^2}\right)_{|A|=0}\kern-1em|A|^2+\ldots\; .
\end{equation}
Going over to the operator equation for the amplitude $A$ by the
substitution $\Delta\omega\rightleftarrows\I\partial/\partial \tau$,
$\Delta k\rightleftarrows-\I\partial/\partial \xi$
and omitting the high-order terms, we get
\begin{equation}\label{eq:NLS}
\I (A_\tau + a_1 A_\xi) - a_2 A_{\xi\xi} + a_{0,\,0,\,0}A|A|^2 = 0,
\end{equation}
where $a_1=(\partial\omega/\partial k)_{k=k_0}$,
$a_2=\frac{1}{2}(\partial^2\omega/\partial k^2)_{k=k_0}$,
$a_{0,\,0,\,0}=(\partial\omega/\partial |A|^2)_{|A|=0}$, and the
variables $\tau$ and $\xi$ stand for some conventional time and
coordinate with respect to which the amplitude of wave train
envelope exhibits slow variations.

Equation \refe{eq:NLS} is called the nonlinear Schr\"{o}dinger
equation (NLSE). It is met, in particular, in various problems of
nonlinear optics, plasma physics, and hydrodynamics, and it admits
solutions in the form of solitons \cite[Sec. 8]{Dodd_solitons}. NLSE
takes into account the second-order dispersion effects (term
$A_{\xi\xi}$) and the phase self-modulation (term $A|A|^2$). These
terms are sufficient to describe the propagation of picosecond
pulses in optical fibers \cite{Ruan2005}. Since solitons are formed
by the balance of dispersion and nonlinearity, a lesser pulse width
can be obtained when the carrier wavelength is chosen such that the
coefficient $a_2$ of the dispersion term vanishes. So,
high-order dispersive and nonlinear effects come to the forefront in
femtosecond-pulse problems. These effects are described by
generalized NLSEs (high-order NLSEs) \cite{Ruan2005,Davydova}. In
the case of hydrodynamic wave propagation along a fluid surface,
high-order NLSEs were derived, in particular, in
\cite{Dysthe,Sedl_UJP_2003,Slunyaev,Das}. Another distinctive
feature of wave trains described by NLSE is their instability with
respect to long-wave modulations (modulational instability) at
$a_2\, a_{0,\,0,\,0} < 0$ \cite[p.~640]{encyclop}. To determine the
modulational instability conditions at $a_{0,\,0,\,0}=0$, high-order
terms are to be taken into account in Eq. \refe{eq:NLS}.

There are several methods which allow the original equations of one
or other physical process to be reduced to model evolution equations
of NLSE type. They include the multiple scale method
\cite{Dodd_solitons,Slunyaev,lukom2001}, Hamiltonian formalism
(Zakharov method)
\cite{zakharov1966,YuenLake:review_rus,Sedl_UJP_2007}, variational
method \cite{Whitham}, and reductive perturbation methods
\cite{taniuti1969,demiray2005}. In the multiple scale method, an
unknown function $u(x,\,t)$ of coordinate and time is looked for in
the form of asymptotic expansion in powers of a small nonlinearity
parameter~$\epsilon$:
\begin{equation}
\label{eq:perturbation} u(x,\,t)=\sum_{n=1}^{\infty}\epsilon^n
u^{(n)}(x,\,t).
\end{equation}
The wave motion is classified into slow one and fast one by
introducing different time scales and different spatial scales:
\[
T_n \equiv \mu^n t,\quad X_n \equiv \mu^n x.
\]
The derivatives with respect to time and coordinate are expanded
into the following series:
\begin{equation}\label{eq:der_expansion}
\frac{\partial}{\partial t} = \sum_{n=0}^{\infty}\mu^n
\frac{\partial}{\partial T_n},\qquad \frac{\partial}{\partial x} =
\sum_{n=0}^{\infty}\mu^n \frac{\partial}{\partial X_n},
\end{equation}
the times $T_n$ and coordinates $X_n$ being assumed to be
independent variables. A principal drawback of this method lies in the
fact that the parameters $\epsilon$ and $\mu$ with different
physical meanings (the former characterizing the smallness of
nonlinearity, and the latter describing the slowness of temporal and
spatial variations) are tentatively taken equal: $\epsilon=\mu$.
This admission produces the so-called secular terms in the equations
for $u^{(n)}(x,\,t)$. Such terms, which infinitely grow with time,
are eliminated in each new order of $\epsilon$ by an appropriate
choice of free parameters emerging in the solutions of the linear
inhomogeneous wave equations derived from the original nonlinear
equations for the function $u(x,\,t)$. The procedure is very
awkward, and it is difficult to formulate in algorithmic form. In
Zakharov's method, the problem is reduced to an integral equation in
the Fourier space, and the corresponding solutions should be
transformed back to the physical space with the use of the inverse
Fourier transformation. Again, the procedure is quite laborious. The same
remarks can be made regarding all other methods mentioned above.

In work \cite{lukom1995}, V.P.~Lukomsky proposed an idea of
constructing a perturbation procedure free of secular terms. It was
used to derive a generalized NLSE for the modulations of gravity
waves on deep water. The method allows the original system of
nonlinear equations to be reduced to a model equation for the pulse
envelope in the form of asymptotic expansion of the time derivative
in terms of two independent parameters which characterize the
smallness of amplitudes ($\varepsilon$) and the slowness of their
spatial variations ($\mu$). In this paper, we present a general
realization of this two-parameter procedure by the example of the
reduction of the Klein--Gordon equation to a generalized NLSE. Our
technique allows the coefficients of the generalized NLSE to be
calculated in arbitrary order of $\varepsilon$ (high-order nonlinear
terms) and $\mu$ (high-order dispersive terms) as well as any their
combination (nonlinear-dispersive terms).

Consider some wave process described by the \mbox{$(1+1)$} Klein--Gordon
equation with arbitrary polynomial nonlinearity:
\begin{equation}\label{KG}
u_{tt}-c^2u_{xx}+\sum_{p=1}^P\alpha_p u^p=0.
\end{equation}
Here $u$ is an unknown twice differentiable function of the wave
process, $0 < t < \infty$ is time, $-\infty < x < \infty$ is
coordinate, $c$ and $\alpha_p$ are arbitrary real constants
($\alpha_1\neq0$), and $P$ is an arbitrary positive integer. Let the
initial condition at $t = 0$ have the form
$u(x,\,0)=Q(x)\bigl(\exp(\I k x)+\exp(-\I k x)\bigr)$,
$u_t(x,\,0)=P(x)\bigl(\exp(\I k x)+\exp(-\I k x)\bigr)$, where $k$
is the carrier wave number.

The Klein--Gordon equation arises in the field theory,
elementary particle physics, crystal dislocation models, etc.
\cite{Dodd_solitons}. When $\alpha_{2p+1}=(-1)^p/(2p+1)!$,
$\alpha_{2p}=0$, $P=\infty$, Eq.~\refe{KG} is called the sin-Gordon
equation, and it is used to model the dynamics of dislocations in
crystals, self-induced transparency in nonlinear optics, spin waves
in fluid helium, propagation of fluxons in long Josephson
(superconductive) junctions, and dynamics of domain walls in
ferromagnetics \cite[p.~840]{encyclop}.

\section{Spectral Representation}

We look for a solution to Eq.~\refe{KG} in the form of truncated
Fourier series with variable coefficients:
\begin{equation}\label{u}
u(x,\,t)=\sum_{n=-N_u}^{N_u}u_n(x,\,t)\E^{\I n(\omega t-kx)},\quad u_{-n}\equiv u_n^*,
\end{equation}
where $\omega$ is the wave-train carrier frequency, $N_u + 1$ is the
number of harmonics taken into consideration, and $^*$ stands for
complex conjugate. The same series can be written for all integer
powers of the function $u$:
\begin{equation}\label{up}
u^p(x,\,t)=\sum_{n=-p N_u}^{p N_u}(u^p)_n(x,\,t)\E^{\I n(\omega t-kx)},\quad p=\overline{2,\,P},
\end{equation}
where $(u^p)_{-n}\equiv (u^p)_n^*$. The coefficients $(u^p)_n$ can
be expressed recurrently in terms of the coefficients $u_n$
\cite[p.~30]{book}:
\[
(u^p)_n=\sum_{n_1=\max(-N_u,\,n-(p-1)N_u)}^{\min(N_u,\,n+(p-1)N_u)} u_{n_1}(u^{p-1})_{n-n_1}.
\]
The corresponding expansions of the derivatives are
\[
u_{tt}(x,\,t)=\!\!\!\!\sum_{n=-N_u}^{N_u}\bigl((u_n)_{tt}+\,2\I n
\omega(u_n)_t-\]
\begin{equation}\label{utt}
-\,n^2\omega^2 u_n\bigr)\E^{\I n(\omega t-kx)},
\end{equation}
\[
u_{xx}(x,\,t)=\!\!\!\!\sum_{n=-N_u}^{N_u}\bigl((u_n)_{xx}-\,2\I n
k(u_n)_x-\]
\begin{equation}
-\,n^2 k^2 u_n\bigr)\E^{\I n(\omega t-kx)}.\label{uxx}
\end{equation}
Substituting \refe{u}--\refe{uxx} in \refe{KG} and equating the
coefficients at the like powers of the exponent $\exp(\I (\omega
t-kx))$, we obtain a system of nonlinear differential equations for
the coefficients $u_n(x,\,t)$ ($n=\overline{0,\,N_u}$):
\[
(u_n)_{tt}-c^2(u_n)_{xx}+2\I n\bigl(\omega(u_n)_t +
c^2k(u_n)_x\bigr)+\]
\begin{equation}\label{eq:un}
+ (n^2c^2 k^2-n^2\omega^2+\alpha_1)u_n + \sum_{p=2}^P\alpha_p
(u^p)_n=0.
\end{equation}
Linearization of these equations at $n=1$ gives the dispersion
relation in the linear approximation:
\begin{equation}\label{eq:dispersion_linear}
\omega^2=\alpha_1+c^2 k^2.
\end{equation}

\section{Two-Parameter Expansions for Narrow-Band Wave Trains}
Generally, the system of equations \refe{eq:un} is by no means more
simple than original equation \refe{KG}. It can be simplified if
solutions are looked for in a class of functions with narrow
spectrum, $|\Delta k| \ll k$ ({\em quasi-monochromaticity}
condition). In this case, the problem has a formal small parameter
$\mu \sim|\Delta k|\bigl/k$, and the coefficients $u_n(x,\,t)$ can
be regarded as slow functions of $x$ and $t$. Let us introduce a
slow coordinate $\xi=\mu x$ and go over to the variables
$u_n=u_n(\mu x,\, t)$.

When there are no resonances between higher harmonics, the
amplitudes of Fourier coefficients decrease with increasing number
({\em quasi-harmonicity} condition):
\begin{equation}\label{eq:harmonicity}
u_n\sim\varepsilon^n A,\;\; n\geqslant1,\quad u_0\sim\varepsilon^2 A,\quad\varepsilon < 1,
\end{equation}
where $u_1\equiv\varepsilon A$. The parameter $\varepsilon$ can be
chosen as a second formal parameter, which is independent of the
dispersion parameter $\mu$ in the general case. The use of two
independent formal parameters is a distinctive feature of our
approach as compared to other perturbative methods (e.g., multiple
scale method), where these parameters are not distinguished
($\varepsilon=\mu$). When these incomparable parameters are set
equal, a perturbative procedure produces non-physical secular terms.

In contrast to perturbative methods which use the expansions of form
\refe{eq:perturbation} and \refe{eq:der_expansion} to reduce
Eqs.~\refe{eq:un} to evolution equations of NLSE type \refe{eq:NLS},
we immediately start from the most general explicit form of such an
evolution equation. To this end, the time derivative $(u_1)_t\equiv
\varepsilon A_t$  should be expressed in terms of the derivatives
$(u_1)_{nx}\equiv \varepsilon\mu^n A_{n\xi}$ with respect to
coordinate (designation $A_{n\xi}$ means the \mbox{$n$-th} derivative with
respect to $\xi$) and all possible combinations of nonlinear terms
$\varepsilon^{2n+1}A^{(n+1)}(A^*)^n$. Hence, the derivative $A_t$
can be written as the following asymptotic expansion in terms of
parameters $\varepsilon$ and $\mu$:
\[
A_t = \I \sum_{n_0=0}^\infty (\I\mu)^{n_0}\Bigl(a_{n_0}
A_{n_0\xi}+\]
\[+\varepsilon^2\!\!\sum_{n_1=0}^{n_0}\sum_{n_2=0}^{n_1}a_{n_0-n_1,\,n_1-n_2,\,n_2}A_{(n_0-n_1)\xi}\times\]
\begin{equation}\label{eq:at}
\times A_{(n_1-n_2)\xi}A_{n_2\xi}^*+O(\varepsilon^4)\Bigr).
\end{equation}
Expression \refe{eq:at} is the general form of the evolution
equation for the complex amplitude $A$ of the first harmonic. The
unknown coefficients $a_{n}\_$ can be determined from
Eqs.~\refe{eq:un}. To this end, the amplitudes of all other harmonics ($u_0$, $u_2$, $u_3,\,\ldots$) are expanded in terms of the amplitude of the first harmonic $A$ in the same manner as it is done in expansion \refe{eq:at}: 
\[u_0 = \varepsilon^2\sum_{n_0=0}^\infty
(\I\mu)^{n_0}\Bigl(\sum_{n_1=0}^{n_0}b^{(0)}_{n_0-n_1,\,n_1}A_{(n_0-n_1)\xi}A_{n_1\xi}^*+\]
\[+
\varepsilon^2\sum_{n_1=0}^{n_0}\sum_{n_2=0}^{n_1}\sum_{n_3=0}^{n_2}b^{(0)}_{n_0-n_1,\,n_1-n_2,\,n_2-n_3,\,n_3}\times\]
\begin{equation}
    \times A_{(n_0-n_1)\xi}A_{(n_1-n_2)\xi}A_{(n_2-n_3)\xi}^*A_{n_3\xi}^*+O(\varepsilon^4)\Bigr),
\end{equation}
\[
u_2 = \varepsilon^2\sum_{n_0=0}^\infty
(\I\mu)^{n_0}\Bigl(\sum_{n_1=0}^{n_0}b^{(2)}_{n_0-n_1,\,n_1}A_{(n_0-n_1)\xi}A_{n_1\xi}\,+\]
\[+
\varepsilon^2\sum_{n_1=0}^{n_0}\sum_{n_2=0}^{n_1}\sum_{n_3=0}^{n_2}b^{(2)}_{n_0-n_1,\,n_1-n_2,\,n_2-n_3,\,n_3}\times\]
\begin{equation}
\times
A_{(n_0-n_1)\xi}A_{(n_1-n_2)\xi}A_{(n_2-n_3)\xi}A_{n_3\xi}^*+O(\varepsilon^4)\Bigr),
\end{equation}
\[u_3= \varepsilon^3\sum_{n_0=0}^\infty
(\I\mu)^{n_0}\Bigl(\sum_{n_1=0}^{n_0}\sum_{n_2=0}^{n_1}b^{(3)}_{n_0-n_1,\,n_1-n_2,\,n_2}\times\]
\[\times A_{(n_0-n_1)\xi}A_{(n_1-n_2)\xi}A_{n_2\xi}\,+\]
\[+
\varepsilon^2\sum_{n_1=0}^{n_0}\sum_{n_2=0}^{n_1}\sum_{n_3=0}^{n_2}\sum_{n_4=0}^{n_3}b^{(3)}_{n_0-n_1,\,n_1-n_2,\,n_2-n_3\,n_3-n_4,\,n_4}\times\]
\begin{equation}\label{eq:u3}
\times
A_{(n_0-n_1)\xi}A_{(n_1-n_2)\xi}A_{(n_2-n_3)\xi}A_{(n_3-n_4)\xi}A_{n_4\xi}^*+O(\varepsilon^4)\Bigr),
\end{equation}\vspace*{-7mm}
\[\ldots\;\;.\]
The unknown coefficients $b^{(n)}_{n\_}$ are found along with the
coefficients $a_{n}\_$ from the system of equations \refe{eq:un}
by substituting expressions \refe{eq:at}--\refe{eq:u3} and
equating the coefficients at the like powers of the products
$\varepsilon^{k}\mu^m$ in different combinations $(A\_\ldots
A^*\_\ldots)$ to zero. In its essence, this procedure is similar
to the method of undetermined coefficients. The coefficient
calculation order and the general form of the expansions for $A_t$
and $u_n$ in arbitrary order of $\varepsilon$ are given in
Appendix. The use of two parameters in ansatz \refe{eq:at} is of
key importance for the coefficient calculation procedure, since
the expansions could not be split into linear-independent terms at
$\varepsilon = \mu$.

Note that our two-parameter approach has the same limitations in
regard to convergence issues as other perturbative methods do. The
convergence can get broken in the presence of resonances between
harmonics, when quasi-harmonicity condition \refe{eq:harmonicity} is
violated. Some questions related to the convergence of asymptotic
expansions for the solutions of differential equations were
considered, in particular, in our works \cite{uniform1,uniform2}.
Expansion \refe{eq:at} cannot be used either for wide-band wave
trains with $\Delta k\sim k$.

It should also be noted that the reduction of Eq.~\refe{KG} with the
second time derivative to Eq.~\refe{eq:at} with the first time
derivative puts a constraint on the initial condition for $u_t$. In
this case, $u_t(x,0)$ is a function of $u(x,0)$ defined by formula
\refe{eq:at}.

\section{High-Order Nonlinear Schr\"{o}dinger Equation}
The two-parameter expansions were programmed in symbolic form for an
arbitrary order of $\mu$ and $\varepsilon$. The evolution equation
for the complex amplitude of the first harmonic is
\[
A_t = \I\Bigl((\I\mu) a_1A_\xi + (\I\mu)^2 a_2 A_{\xi\xi} +
\]
\[+ (\I\mu)^3 a_3 A_{\xi\xi\xi} + (\I\mu)^4 a_4 A_{\xi\xi\xi\xi} +
O(\mu^5)+\]
\[+ \varepsilon^2\Bigl[a_{0,\,0,\,0}A|A|^2 +  (\I\mu)\phantom{^2}\bigl(a_{1,\,0,\,0}A_\xi |A|^2 +
a_{0,\,0,\,1}A^2 A_\xi^*\bigr)+
\]
\[+ (\I\mu)^2\bigl(a_{2,\,0,\,0}A_{\xi\xi}|A|^2 + a_{1,\,1,\,0}A_{\xi}^2 A^* +
\]
\[+ a_{1,\,0,\,1}|A_\xi|^2 A + a_{0,\,0,\,2}A^2 A_{\xi\xi}^*\bigr) + O(\mu^3)\Bigr]+
\]
\[+ \varepsilon^4\Bigl[a_{0,\,0,\,0,\,0,\,0}A|A|^4 + \]
\[+ (\I\mu)\phantom{^2}\bigl(a_{1,\,0,\,0,\,0,\,0}A_\xi |A|^4 + a_{0,\,0,\,0,\,1,\,0}A^2|A|^2 A_\xi^*\bigr)+
\]
\[+ (\I\mu)^2\bigl(a_{2,\,0,\,0,\,0,\,0}A_{\xi\xi}|A|^4 + a_{1,\,1,\,0,\,0,\,0}A_{\xi}^2 |A|^2 A^* +
\]
\[+ a_{1,\,0,\,0,\,1,\,0}|A_\xi|^2 A |A|^2 + a_{0,\,0,\,0,\,2,\,0}A^2 |A|^2 A_{\xi\xi}^*
+\]
\begin{equation}
\label{GNLS}
 + a_{0,\,0,\,0,\,1,\,1}A^3 A_\xi^2\bigr) + O(\mu^3)\Bigr]+
O(\varepsilon^6)\Bigr).
\end{equation}
In each term of this equation, the power of the formal parameter
$\mu$ points to the overall order of the derivatives with respect to
$\xi$, and the power of the formal parameter $\varepsilon$ points to
the nonlinearity order. These parameters disappear after going back
to the original variables $u_1=\varepsilon A$ and $x=\xi/\mu$.
Taking into account dispersion relation \refe{eq:dispersion_linear},
the coefficients $a\_$ can be written as (at $P=5$)
\[
a_0 = 0,\quad a_1 = \frac{c^2 k}{\omega },\quad a_2 = \frac{c^2
\alpha_1}{2\omega^3},\quad a_3 = -\frac{c^4 k \alpha_1}{2 \omega^5},
\]
\[
a_4 = \frac{\alpha_1 c^4 (4 c^2 k^2 - \alpha_1)}{8 \omega^7},\quad
a_n = \frac{1}{n!}\frac{\D^n\omega}{\D k^n};
\]
\[
a_{0,\,0,\,0} = \frac{3\alpha _3}{2 \omega }-\frac{5 \alpha _2^2}{3
\omega  \alpha _1},
\]
\[
a_{1,\,0,\,0} = 2a_{0,\,0,\,1} = \frac{c^2 k}{\omega
^3}\biggl(\frac{10 \alpha _2^2}{3 \alpha _1}-3 \alpha _3\biggr);
\]
\[
a_{2,\,0,\,0} = \frac{c^2}{18\alpha_1\omega^5} \Bigl( 2c^2
k^2\bigl(27 \alpha _1 \alpha _3-14 \alpha _2^2\bigr)
-\]
\[-
\alpha_1\bigr(27 \alpha _1 \alpha _3 - 62 \alpha
_2^2\bigl)\Bigr),\]
\[
a_{1,\,1,\,0} = \frac{c^2}{36\alpha_1\omega^5} \Bigl( 4c^2
k^2\bigl(27 \alpha _1 \alpha _3-28 \alpha _2^2\bigr) -
\]
\[
-\alpha_1\bigr(27 \alpha _1 \alpha _3 - 38 \alpha
_2^2\bigl)\Bigr),\]
\[a_{1,\,0,\,1} = \frac{c^2}{6\alpha_1\omega^5} \Bigl(
4c^2 k^2\bigl(9 \alpha _1 \alpha _3-4 \alpha _2^2\bigr) -
\]
\[-\alpha_1\bigr(9 \alpha _1 \alpha _3 - 34 \alpha
_2^2\bigl)\Bigr),\]
\[
a_{0,\,0,\,2} = \frac{c^2}{6\alpha_1\omega^5} \Bigl( c^2 k^2\bigl(9
\alpha _1 \alpha _3+2 \alpha _2^2\bigr) + 12 \alpha_1 \alpha
_2^2\Bigr);
\]
\[
a_{0,\,0,\,0,\,0,\,0}=\frac{1}{\omega}\Bigl(-\frac{335
\alpha_2^4}{108\alpha _1^3}-\frac{25 \alpha _2^4}{18 \omega ^2
\alpha _1^2}+\frac{143 \alpha _2^2 \alpha _3}{12 \alpha
_1^2}+\]
\[+\frac{5 \alpha _2^2 \alpha _3}{2 \omega ^2 \alpha _1}-\frac{9 \alpha
_3^2}{8 \omega ^2}+\frac{3 \alpha _3^2}{16 \alpha _1}-\frac{14
\alpha _2 \alpha _4}{\alpha _1}+5 \alpha _5\Bigr),\]
\[
a_{1,\,0,\,0,\,0,\,0}=\frac{c^2 k}{\omega^3}\Bigl(\frac{925 \alpha
_2^4}{108 \alpha _1^3}+\frac{275 \alpha _2^4}{18 \omega ^2 \alpha
_1^2}-\frac{421 \alpha _2^2 \alpha _3}{12 \alpha
_1^2}-\]
\[-
\frac{55 \alpha _2^2 \alpha _3}{2 \omega ^2 \alpha _1}+\frac{99
\alpha _3^2}{8 \omega ^2}-\frac{9 \alpha _3^2}{16 \alpha
_1}+\frac{42 \alpha _2 \alpha _4}{\alpha _1}-15 \alpha _5\Bigr),
\]
\[a_{0,\,0,\,0,\,1,\,0}=\frac{c^2 k}{\omega^3}\Bigl(\frac{295 \alpha _2^4}{54 \alpha _1^3}+\frac{100 \alpha _2^4}{9 \omega ^2 \alpha _1^2}-\frac{139 \alpha _2^2 \alpha _3}{6 \alpha
_1^2}-\]
\[-\frac{20 \alpha _2^2 \alpha _3}{\omega ^2 \alpha _1}+\frac{9 \alpha _3^2}{\omega ^2}-\frac{3 \alpha _3^2}{8 \alpha _1}+\frac{28 \alpha _2 \alpha _4}{\alpha _1}-10 \alpha _5\Bigr).
\]
The expressions for the subsequent coefficients $a\_$ are too long
to be presented in explicit form.

The complex amplitudes of other harmonics are found from the
relations
\[
u_0 = \varepsilon^2\Bigl(\Bigl[b^{(0)}_{0,\,0}|A|^2 +
(\I\mu)\bigl(b^{(0)}_{1,\,0}A_\xi A^* + b^{(0)}_{0,\,1}A
A_\xi^*\bigr)+ \]
\[+ (\I\mu)^2\bigl(b^{(0)}_{2,\,0}A_{\xi\xi}A^* + b^{(0)}_{1,\,1}A_{\xi}A_{\xi}^* + b^{(0)}_{0,\,2} A A_{\xi\xi}^*\bigr) +
\]
\[+ O(\mu^3)\Bigr]+\varepsilon^2\Bigl[b^{(0)}_{0,\,0,\,0,\,0}|A|^4+
\]
\[+ (\I\mu)\phantom{^2}\bigl(b^{(0)}_{1,\,0,\,0,\,0}A_\xi |A|^2A^* + b^{(0)}_{0,\,0,\,1,\,0}A |A|^2 A_\xi^*\bigr)+
\]
\[+ (\I\mu)^2\bigl(b^{(0)}_{2,\,0,\,0,\,0}A_{\xi\xi}|A|^2A^* + b^{(0)}_{1,\,1,\,0,\,0}A_{\xi}^2 {A^*}^2 +
\]
\[+ b^{(0)}_{1,\,0,\,1,\,0}|A_\xi|^2 |A|^2 + b^{(0)}_{0,\,0,\,2,\,0}A |A|^2 A_{\xi\xi}^*
+\]
\[+ b^{(0)}_{0,\,0,\,1,\,1}A^2 {A_\xi^*}^2\bigr) +
O(\mu^3)\Bigr]+ O(\varepsilon^4)\Bigr);\]
\[u_2 = \varepsilon^2\Bigl(\Bigl[b^{(2)}_{0,\,0}A^2 + (\I\mu)b^{(2)}_{1,\,0}A_\xi A+
\]
\[+ (\I\mu)^2\bigl(b^{(2)}_{2,\,0}A_{\xi\xi}A + b^{(2)}_{1,\,1}A_{\xi}^2\bigr) + O(\mu^3)\Bigr]+
\]
\[+ \varepsilon^2\Bigl[b^{(2)}_{0,\,0,\,0,\,0}A^2|A|^2 + \]
\[+ (\I\mu)\phantom{^2}\bigl(b^{(2)}_{1,\,0,\,0,\,0}A_\xi A |A|^2 + b^{(2)}_{0,\,0,\,0,\,1}A^3 A_\xi^*\bigr)+
\]
\[+ (\I\mu)^2\bigl(b^{(2)}_{2,\,0,\,0,\,0}A_{\xi\xi}A|A|^2 + b^{(2)}_{1,\,1,\,0,\,0}A_{\xi}^2 |A|^2 +
\]
\[+ b^{(2)}_{1,\,0,\,0,\,1}|A_\xi|^2 A^2 + b^{(2)}_{0,\,0,\,0,\,2}A^3 A_{\xi\xi}^* \bigr) + O(\mu^3)\Bigr]+
 O(\varepsilon^4)\Bigr);\]
\[
u_3 = \varepsilon^3\Bigl(\Bigl[b^{(3)}_{0,\,0,\,0}A^3 +
(\I\mu)b^{(3)}_{1,\,0,\,0}A_\xi A^2+ \]
\[+ (\I\mu)^2 \bigl(b^{(3)}_{2,\,0,\,0}A_{\xi\xi}A^2 + b^{(3)}_{1,\,1,\,0}A_{\xi}^2A\bigr) +
O(\mu^3)\Bigr]+ O(\varepsilon^2)\Bigr).\] At $P=5$, the
coefficients of the expansions are
\[
b^{(0)}_{0,\,0} = -\frac{2\alpha_2}{\alpha_1};\quad b^{(0)}_{1,\,0}
= b^{(0)}_{0,\,1} = 0;
\]
\[
b^{(0)}_{2,\,0} = \frac{1}{2}\,b^{(0)}_{1,\,1} = b^{(0)}_{0,\,2} =
\frac{2c^2\alpha_2}{\omega^2\alpha_1};
\]
\[
b^{(0)}_{0,\,0,\,0,\,0} = -\frac{38 \alpha _2^3}{9 \alpha
_1^3}+\frac{10 \alpha _2 \alpha _3}{\alpha _1^2}-\frac{6 \alpha
_4}{\alpha _1};
\]
\[
b^{(0)}_{1,\,0,\,0,\,0} = b^{(0)}_{0,\,0,\,1,\,0} = 0;
\]
\[
b^{(0)}_{2,\,0,\,0,\,0}=b^{(0)}_{0,\,0,\,2,\,0}=\]
\[=\frac{c^2}{27\alpha_1^3\omega^4}\Bigl(-4 c^2 k ^2 \left(23 \alpha _2^3+90 \alpha _1 \alpha _2 \alpha _3-81 \alpha _1^2 \alpha
_4\right)+\]
\[+\alpha _1 \left(538 \alpha _2^3-927 \alpha _1 \alpha
_2 \alpha _3+324 \alpha _1^2 \alpha _4\right)\Bigr),
\]
\[
b^{(0)}_{1,\,1,\,0,\,0}=b^{(0)}_{0,\,0,\,1,\,1}=\]
\[=\frac{c^2}{27\alpha_1^3\omega^4}\Bigl(-4 c^2 k ^2 \left(79 \alpha _2^3+18 \alpha _1 \alpha _2 \alpha _3-81 \alpha _1^2 \alpha
_4\right)+\]
\[+\alpha _1 \left(314 \alpha _2^3-639
\alpha _1 \alpha _2 \alpha _3+324 \alpha _1^2 \alpha
_4\right)\Bigr),\]
\[
b^{(0)}_{1,\,0,\,1,\,0}=\]
\[=\frac{4c^2}{9\alpha_1^3\omega^4}\Bigl(-4 c^2 k ^2 \left(17 \alpha _2^3+18 \alpha _1 \alpha _2 \alpha _3-27 \alpha _1^2 \alpha
_4\right)+\]
\[+\alpha _1 \left(142 \alpha _2^3-261 \alpha _1 \alpha
_2 \alpha _3+108 \alpha _1^2 \alpha
_4\right)\Bigr);\]
\[
b^{(2)}_{0,\,0} = \frac{\alpha_2}{3\alpha_1};\quad b^{(2)}_{1,\,0} =
0;\quad b^{(2)}_{2,\,0} = -b^{(2)}_{1,\,1} =
-\frac{2c^2\alpha_2}{9\omega^2\alpha_1};
\]
\[
b^{(2)}_{0,\,0,\,0,\,0} = \frac{59 \alpha _2^3}{54 \alpha
_1^3}-\frac{31 \alpha _2 \alpha _3}{12 \alpha _1^2}+\frac{4 \alpha
_4}{3 \alpha _1};
\]
\[
b^{(2)}_{1,\,0,\,0,\,0} = b^{(2)}_{0,\,0,\,0,\,1}= \frac{2c^2 k
\alpha_2}{27\omega^2\alpha_1^3}\bigl(9\alpha_1\alpha_3-10\alpha_2^2\bigr);
\]
\[
b^{(2)}_{2,\,0,\,0,\,0}=-\frac{c^2}{1296\alpha_1^3\omega^4}\Bigl(c^2
k ^2 \bigl(2606 \alpha _2^3-5283 \alpha _1 \alpha _2 \alpha
_3+\]
\[+1728 \alpha _1^2 \alpha _4\bigr)+\alpha _1 \left(5006 \alpha
_2^3-7443 \alpha _1 \alpha _2 \alpha _3+1728 \alpha _1^2 \alpha
_4\right)\Bigr),\]
\[
b^{(2)}_{1,\,1,\,0,\,0}=\frac{c^2}{1296\alpha_1^3\omega^4}\Bigl(c^2
k ^2 \bigl(3574 \alpha _2^3-7191 \alpha _1 \alpha _2 \alpha
_3+\]
\[+3456 \alpha _1^2 \alpha _4\bigr)+\alpha _1 \left(3094 \alpha
_2^3-6759 \alpha _1 \alpha _2 \alpha _3+3456 \alpha _1^2 \alpha
_4\right)\Bigr),\]
\[
b^{(2)}_{1,\,0,\,0,\,1}=\frac{c^2}{162\alpha_1^3\omega^4}\Bigl(c^2
k ^2 \bigl(122 \alpha _2^3-369 \alpha _1 \alpha _2 \alpha
_3+\]
\[+432 \alpha _1^2 \alpha _4\bigr)+\alpha _1 \left(-358 \alpha
_2^3+63 \alpha _1 \alpha _2 \alpha _3+432 \alpha _1^2 \alpha
_4\right)\Bigr),\]
\[
b^{(2)}_{0,\,0,\,0,\,2}=\frac{c^2}{324\alpha_1^3\omega^4}\Bigl(c^2
k ^2 \bigl(2 \alpha _2^3-261 \alpha _1 \alpha _2 \alpha
_3+\]
\[+432 \alpha _1^2 \alpha _4\bigr)+\alpha _1 \left(-238 \alpha
_2^3-45 \alpha _1 \alpha _2 \alpha _3+432 \alpha _1^2 \alpha
_4\right)\Bigr);\]
\[
b^{(3)}_{0,\,0,\,0} = \frac{\alpha _2^2}{12 \alpha
_1^2}+\frac{\alpha _3}{8 \alpha _1};\quad b^{(3)}_{1,\,0,\,0} =
0;\]
\[b^{(3)}_{2,\,0,\,0} = -b^{(3)}_{1,\,1,\,0}= -\frac{c^2}{288\alpha
_1^2\omega^2}\bigl(27\alpha_1\alpha_3+34\alpha_2^2\bigr).\]

Some of these coefficients were derived earlier in \cite{lukom2001}
by the multiple scale method. The expressions presented in
\cite{lukom2001} are in agreement with those obtained here (except
for several misprints and typographic errors).

\section{The Effect of High-Order Dispersive Terms}

Let us illustrate the evolution of a wave train envelope described
by the equation of form \refe{eq:at}. To this end, we rewrite
original equation \refe{KG} in dimensionless variables
$\widetilde{x}\equiv kx$ and $\widetilde{t}\equiv ckt$:
\begin{equation}\label{KG_dim}
\widetilde{u}_{\widetilde{t}\widetilde{t}}-\widetilde{u}_{\widetilde{x}\widetilde{x}}+\sum_{p=1}^P\widetilde{\alpha}_p
\widetilde{u}^p=0,\quad \widetilde{u} = \frac{u}{U_0},\quad
\widetilde{\alpha}_p=\frac{\alpha_p U_0^{p-1}}{(ck)^2}.
\end{equation}
In this case, we have $\widetilde{c}=1$, $\widetilde{k}=1$,
$\widetilde{\omega}^2=\widetilde{\alpha}_1+1$, and $U_0$ is a
typical amplitude of the function $u$.

As an example, let us consider the case $P=3$ with
$\widetilde{\alpha}_1=1$, $\widetilde{\alpha}_2=0$, and
$\widetilde{\alpha}_3=-1/6$. The values of these parameters
correspond to the first two terms in the Taylor expansion of the
function $\sin u$. Hereafter, the tildes over the dimensionless
variables are omitted.

The corresponding coefficients of evolution equation \refe{GNLS} are
\[
a_1 = \frac{1}{\sqrt{2}},\quad a_2 = \frac{1}{4\sqrt{2}},\quad a_3 =
-\frac{1}{8 \sqrt{2}},\quad a_4 = \frac{3}{64 \sqrt{2}},
\]
\[
a_{0,\,0,\,0} = -\frac{1}{4\sqrt{2}},\quad a_{1,\,0,\,0} =
\frac{1}{4\sqrt{2}},\quad a_{0,\,0,\,1} = \frac{1}{8\sqrt{2}},
\]
\[
a_{2,\,0,\,0} = a_{0,\,0,\,2} = -\frac{1}{16\sqrt{2}},\;\;
a_{1,\,1,\,0} = a_{1,\,0,\,1} = -\frac{3}{16\sqrt{2}},
\]
\begin{equation}\label{a_n}
a_{0,\,0,\,0,\,0,\,0} = -\frac{1}{96\sqrt{2}},\quad \ldots\;.
\end{equation}

Initially, we retain only those terms in Eq.~\refe{GNLS} whose
overall order of smallness with respect to the parameters
$\varepsilon$ and $\mu$ is no more than two. In this case, we obtain
a classical NLSE:
\begin{equation}\label{NLSE}
(u_1)_t = -a_1 (u_1)_x - \I a_2 (u_1)_{xx} + \I a_{0,\,0,\,0}\,u_1|u_1|^2.
\end{equation}

\begin{center} \noindent \epsfxsize=\columnwidth\epsffile{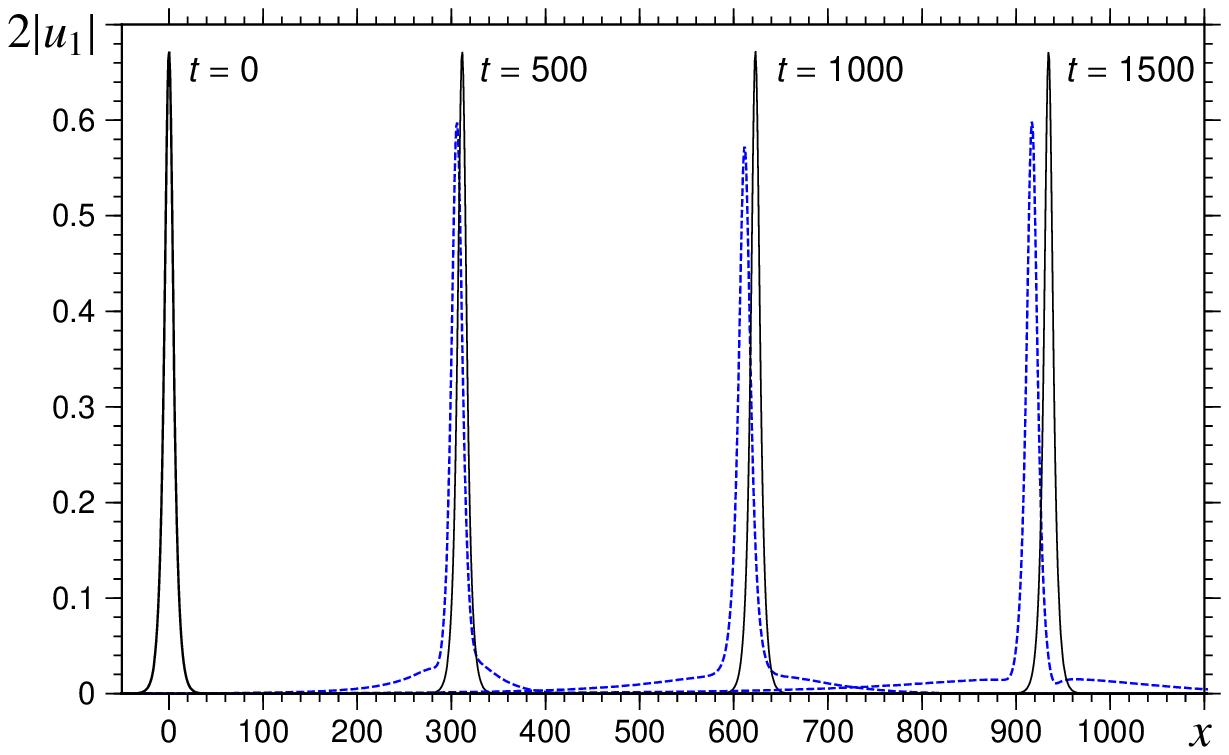}
\end{center}

\noindent{\footnotesize Evolution of the wave train
envelope which is given, at the initial moment $t=0$, by function
\refe{soliton} with parameters $\beta=\zeta=1/10$ and
$\phi_0=x_0=0$. (Solid curve) exact one-soliton solution of NLSE
\refe{NLSE}, (dashed curve) numerical solution of the generalized
NLSE~\refe{GNLSE}}%
\vspace*{3mm}

It has an exact one-soliton solution at $a_2\, a_{0,\,0,\,0} < 0$:
\[u_1(x,\,t)=\beta\left(\frac{2}{|a_{0,\,0,\,0}|}\right)^{1/2}\times\]
\[\times\exp\left(\I\Bigl(\frac{\zeta}{\sqrt{|a_2|}}(x-a_1 t)-s(\zeta^2-\beta^2)t +
\phi_0\Bigr)\right)\times\]
\begin{equation}\label{soliton}
\times\cosh^{-1}\left(\frac{\beta}{\sqrt{|a_2|}}(x-x_0-a_1
t)-2s\beta\zeta t\right).
\end{equation}
Here $s=\mathrm{sign}(a_{0,\,0,\,0})$ and $\beta$, $\zeta$,
$\phi_0$, and $x_0$ are free parameters
\cite{chen1993,zakharov1971}. The corresponding approximate solution
of Eq.~\refe{KG_dim} is
\[u(x,\,t)=u_1(x,\,t)\exp\bigl(\I(\sqrt{2}\,t-x)\bigr)+\]
\begin{equation}
+u_1^*(x,\,t)\exp\bigl(-\I(\sqrt{2}\,t-x)\bigr).
\end{equation}


To analyze the effect of high-order dispersive terms on the shape of
one-soliton solution \refe{soliton}, we consider the generalized
NLSE
\[
(u_1)_t = -a_1 (u_1)_x - \I a_2 (u_1)_{xx} + a_3 (u_1)_{xxx} + \]
\[+ \I a_4 (u_1)_{xxxx} + \I a_{0,\,0,\,0}\,u_1|u_1|^2 -\]
\[- a_{1,\,0,\,0}(u_1)_x |u_1|^2 - a_{0,\,0,\,1}\,u_1^2 (u_1^*)_x -
\]
\[- \I a_{2,\,0,\,0}(u_1)_{xx}|u_1|^2 -\I a_{1,\,1,\,0}(u_1)_{x}^2\, u_1^* -
\]
\[- \I a_{1,\,0,\,1}|(u_1)_x|^2 u_1 -\I a_{0,\,0,\,2}\,u_1^2
(u_1^*)_{xx} + \]
\begin{equation}\label{GNLSE}
+ \I a_{0,\,0,\,0,\,0,\,0}\,u_1|u_1|^4
\end{equation}
with the coefficients defined by \refe{a_n}. The initial condition
is chosen in the form of function \refe{soliton} with
$\beta=\zeta=1/10$ and $\phi_0=x_0=0$. The figure shows
the evolution of such an envelope. To solve Eq.~\refe{GNLSE}
numerically, we used the split-step Fourier method
\cite{Agrawal,Muslu2005}. High-order dispersive terms are seen to
affect the amplitude, shape, and velocity of the soliton solution.

\section{Conclusion}

We described a general method for deriving the evolution
equations for narrow-band wave trains in nonlinear media with
dispersion. The procedure produces no secular terms and can easily
be put in algorithmic form. By using the Klein--Gordon equation
with arbitrary polynomial nonlinearity as an example, we
derived a generalized NLSE whose coefficients can be calculated in
any order with respect to nonlinearity and dispersion. The
equation can be used to investigate the propagation of ultrashort
pulses in optical fibers, to study wave events on a fluid surface,
and to describe the Langmuir waves in hot plasmas. \vskip3mm

I.S. Gandzha thanks the State Fundamental Research Foundation of
Ukraine for the financial support (grant of the President of
Ukraine No.~GP/F26/0056).

\noindent

\section*{Postscript}
The  major  part  of  this  paper  was  written  during  the last
months  of  Vasyl  Petrovich  Lukomsky's  lifetime (January 14, 1942
-- March 31, 2008). The above-described two-parameter method was
developed by him as far back as at the end of the 1980s (some of his
notes regarding NLSE were dated by 1986 and 1987). Regretfully, the
results of those studies were published with delay and only in one
paper of 1995 \cite{lukom1995}. Vasyl Petrovich was attracted by
various fields of physics, he had very broad physical horizons and
often sacrificed the time for paper preparation in favor of
something new. Still unpublished is the paper ``Method of indefinite
coefficients for derivation of evolution equations with higher terms
for nonlinear waves'' written together with Yu.G. Rapoport and
submitted to Physica Scripta in 1999. They derived a high-order NLSE
for describing the propagation of ion sound in non-isothermal
plasmas. Vasyl Petrovich turned back to his method (which can
appropriately be called the Lukomsky method) in 2005 with the aim to
derive a generalized NLSE for the modulations of gravity waves on a
fluid surface. This paper was scheduled as the first out of the
whole series of papers devoted to high-order evolution equations. To
our deep sorrow, these plans were ruined by the fatal malady and
untimely death of Vasyl Petrovich. Only time will show whether the
work which had started can be finished without its inspirer.

Vasyl Petrovich was a kind, sincere, calm, even-tempered, tolerant,
open, and generous man, who was always ready to help. He was
faithful to his life principles and convictions till the end. He
had ingenious non-standard way of thinking and well-trained
intuition. He was a man of word and justice. Blessed memory about
Vasyl Petrovich Lukomsky will always abide in the hearts of those
people who had the honor to be in fellowship and collaboration with
him.

\section*{\footnotesize APPENDIX}
{\footnotesize The general forms of the expansions for $A_t$ and
$u_n$ are
\[
A_t = \I \sum_{n_0=0}^\infty
(\I\mu)^{n_0}\sum_{k=0}^\infty\varepsilon^{2k}(u_1)_{n_0,\,2k+1}\;,\]
\[(u_1)_{n_0,\,2k+1}=\]
\[=\sum_{n_1=0}^{n_0}\sum_{n_2=0}^{n_1}\ldots\sum_{n_{2k}=0}^{n_{2k-1}} a_{n_0-n_1,\,n_1-n_2,\,\ldots,\,n_{2k-1}-n_{2k},\,n_{2k}}\times
\]
\[\times\prod_{i=1}^{k+1}A_{(n_{i-1}-n_i)\xi}\prod_{i=k+2}^{2k+1}A_{(n_{i-1}-n_i)\xi}^*\;,\;\;n_{2k+1}\equiv0;\]
\[
u_n =\varepsilon^n \sum_{n_0=0}^\infty
(\I\mu)^{n_0}\sum_{k=0}^\infty\varepsilon^{2k}(u_n)_{n_0,\,2k+n}\;,\;n>2,\]
\[(u_n)_{n_0,\,2k+n}=\sum_{n_1=0}^{n_0}\sum_{n_2=0}^{n_1}\ldots\]
\[\ldots\sum_{n_{2k+n-1}=0}^{n_{2k+n-2}}
b^{(n)}_{n_0-n_1,\,n_1-n_2,\,\ldots,\,n_{2k+n-2}-n_{2k+n-1},\,n_{2k+n-1}}\times
\]
\[\times\prod_{i=1}^{k+n}A_{(n_{i-1}-n_i)\xi}\prod_{i=k+n+1}^{2k+n}A_{(n_{i-1}-n_i)\xi}^*\,,\;\;n_{2k+n}\equiv0;\]
\[
u_0 =\varepsilon^2 \sum_{n_0=0}^\infty
(\I\mu)^{n_0}\sum_{k=0}^\infty\varepsilon^{2k}(u_0)_{n_0,\,2k+2\;},
\]
\[
(u_0)_{n_0,\,2k+2}=\]
\[=\sum_{n_1=0}^{n_0}\sum_{n_2=0}^{n_1}\ldots\sum_{n_{2k+1}=0}^{n_{2k}}
b^{(0)}_{n_0-n_1,\,n_1-n_2,\,\ldots,\,n_{2k}-n_{2k+1},\,n_{2k+1}}\times\]
\[\times\prod_{i=1}^{k+1}A_{(n_{i-1}-n_i)\xi}\prod_{i=k+2}^{2k+2}A_{(n_{i-1}-n_i)\xi}^*\;,\;\;n_{2k+2}\equiv0.\]
The sums above contain many identical summands (e.g.,
$a_{1,\,0,\,0}A_\xi A A^*$ and $a_{0,\,1,\,0}A A_\xi A^*$). The
repetitions can be eliminated if one limits the summation orders
using the rules $n_{i-2}-n_{i-1} \leqslant n_{i-1}-n_i\leqslant
n_{i}-n_{i+1}$. The corresponding expansions are
\[
(u_1)_{n_0,\,2k+1}=\sum_{n_1=0}^{n_0}\;\sum_{n_2=\max(0,\,2n_1-n_0)}^{n_1}\ldots\]
\[\ldots\sum_{n_i=\max(0,\,2n_{i-1}-n_{i-2})}^{n_{i-1}}\ldots\sum_{n_{k+1}=\max(0,\,2n_{k}-n_{k-1})}^{n_{k}}\]
\[\sum_{n_{k+2}=0}^{[\frac{k-1}{k}n_{k+1}]}\ldots\sum_{n_{i}=\max(0,\,2n_{i-1}-n_{i-2})}^{[\frac{2k+1-i}{2k+2-i}n_{i-1}]}\ldots\]
\[\ldots\sum_{n_{2k}=\max(0,\,2n_{2k-1}-n_{2k-2})}^{[\frac{1}{2}n_{2k-1}]}
a_{n_0-n_1,\,n_1-n_2,\,\ldots,\,n_{2k-1}-n_{2k},\,n_{2k}}\times\]
\[\times\prod_{i=1}^{k+1}A_{(n_{i-1}-n_i)\xi}\prod_{i=k+2}^{2k+1}A_{(n_{i-1}-n_i)\xi}^*\;,\;\;n_{2k+1}\equiv0;\]
\[
(u_{n>1})_{n_0,\,2k+n}|_{k>0}=\sum_{n_1=0}^{n_0}\;\sum_{n_2=\max(0,\,2n_1-n_0)}^{n_1}\ldots\]
\[\ldots\sum_{n_i=\max(0,\,2n_{i-1}-n_{i-2})}^{n_{i-1}}\ldots\kern-0.25cm\sum_{n_{k+n}=\max(0,\,2n_{k+n-1}-n_{k+n-2})}^{n_{k+n-1}}\]
\[\sum_{n_{k+n+1}=0}^{[\frac{k-1}{k}n_{k+n}]}\ldots\sum_{n_{i}=\max(0,\,2n_{i-1}-n_{i-2})}^{[\frac{2k+n-i}{2k+n-i+1}n_{i-1}]}\ldots\]
\[\ldots\sum_{n_{2k+n-1}=\max(0,\,2n_{2k+n-2}-n_{2k+n-3})}^{[\frac{1}{2}n_{2k+n-2}]}
b^{(n)}_{\_}\;\prod_{i=1}^{k+n}A_{\_}\prod_{i=k+n+1}^{2k+n}A_{\_}^*;\]
}

\end{multicols}

\begin{table*}[t]
\noindent{\footnotesize{\bf%
Coefficient calculation order }\vskip1mm \tabcolsep31.5pt

\noindent\begin{tabular}{c c l}
 \hline \multicolumn{1}{c}
{\rule{0pt}{9pt}Iteration number } & \multicolumn{1}{|c}{Number of Eqs.~\refe{eq:un}}& \multicolumn{1}{|c}{Coefficient}\\%
\hline%
\rule{0pt}{9pt}$k=0$ ($\varepsilon^1$) & $n=1$ ($\varepsilon^1$) & $a_{n_0}$\\[2mm]%
$k=1$ ($\varepsilon^3$) & $n=0$ ($\varepsilon^2$) & $b^{(0)}_{n_0-n_1,\,n_1}$\\%
& $n=2$ ($\varepsilon^2$) & $b^{(2)}_{n_0-n_1,\,n_1}$\\%
& $n=1$ ($\varepsilon^3$) & $a_{n_0-n_1,\,n_1-n_2,\,n_2}$\\[2mm]%
$k=2$ ($\varepsilon^5$) & $n=0$ ($\varepsilon^4$) & $b^{(0)}_{n_0-n_1,\,n_1-n_2,\,n_2-n_3,\,n_3}$\\%
& $n=3$ ($\varepsilon^3$) & $b^{(3)}_{n_0-n_1,\,n_1-n_2,\,n_2}$\\%
& $n=2$ ($\varepsilon^4$) & $b^{(2)}_{n_0-n_1,\,n_1-n_2,\,n_2-n_3,\,n_3}$\\%
& $n=1$ ($\varepsilon^5$) & $a_{n_0-n_1,\,n_1-n_2,\,n_2-n_3,\,n_3-n_4,\,n_4}$\\[2mm]%
$k=3$ ($\varepsilon^7$) & $n=0$ ($\varepsilon^6$) & $b^{(0)}_{n_0-n_1,\,n_1-n_2,\,n_2-n_3,\,n_3-n_4,\,n_4-n_5,\,n_5}$\\%
& $n=4$ ($\varepsilon^4$) & $b^{(4)}_{n_0-n_1,\,n_1-n_2,\,n_2-n_3,\,n_3}$\\%
& $n=3$ ($\varepsilon^5$) &
$b^{(3)}_{n_0-n_1,\,n_1-n_2,\,n_2-n_3,\,n_3-n_4,\,n_4}$\\%
& $n=2$ ($\varepsilon^6$) & $b^{(2)}_{n_0-n_1,\,n_1-n_2,\,n_2-n_3,\,n_3-n_4,\,n_4-n_5,\,n_5}$\\%
& $n=1$ ($\varepsilon^7$) & $a_{n_0-n_1,\,n_1-n_2,\,n_2-n_3,\,n_3-n_4,\,n_4-n_5,\,n_5-n_6,\,n_6}$\\[1mm]%
\hline
\end{tabular}\vspace{10mm}
}
\end{table*}

\begin{multicols}{2}

{\footnotesize \noindent\[
(u_{n>1})_{n_0,\,2k+n}|_{k=0}=\sum_{n_{1}=0}^{[\frac{n-1}{n}n_{0}]}\ldots\kern-0.5cm\sum_{n_{i}=\max(0,\,2n_{i-1}-n_{i-2})}^{[\frac{n-i}{n-i+1}n_{i-1}]}\ldots\]
\[\ldots\sum_{n_{n-1}=\max(0,\,2n_{n-2}-n_{n-3})}^{[\frac{1}{2}n_{n-2}]}b^{(n)}_{\_}\prod_{i=1}^{n}A_{\_};\]\vspace*{-3mm}
\[(u_{0})_{n_0,\,2k+2}|_{k>0}=\sum_{n_1=0}^{n_0}\;\sum_{n_2=\max(0,\,2n_1-n_0)}^{n_1}\ldots\]\vspace*{-3mm}
\[\ldots\sum_{n_i=\max(0,\,2n_{i-1}-n_{i-2})}^{n_{i-1}}\ldots\kern-0.25cm\sum_{n_{k+1}=\max(0,\,2n_{k}-n_{k-1})}^{n_{k}}\]\vspace*{-3mm}
\[\sum_{n_{k+2}=0}^{[\frac{k}{k+1}n_{k+1}]}\ldots\sum_{n_{i}=\max(0,\,2n_{i-1}-n_{i-2})}^{[\frac{2k-i+2}{2k-i+3}n_{i-1}]}\ldots\]\vspace*{-3mm}
\[\ldots\sum_{n_{2k+1}=\max(0,\,2n_{2k}-n_{2k-1})}^{[\frac{1}{2}n_{2k}]}\kern-0.5cm
b^{(0)}_{\_}\;\prod_{i=1}^{k+1}A_{\_}\prod_{i=k+2}^{2k+2}A_{\_}^*;\]\vspace*{-3mm}
\[
(u_{0})_{n_0,\,2k+2}|_{k=0}=\sum_{n_{1}=0}^{n_{0}}b^{(0)}_{n_0-n_1,\,n_1}A_{(n_0-n_1)\xi}A_{n_1\xi}^*.
\]

The table gives the order of calculation of the coefficients
$b^{(n)}_{\_}$ and $a_{\_}$. }

\vspace{10mm}

\rezume{ДВОПАРАМЕТРИЧНИЙ МЕТОД ДЛЯ ОПИСУ \\НЕЛІНІЙНОЇ ЕВОЛЮЦІЇ
СПЕКТРАЛЬНО \\ВУЗЬКИХ ХВИЛЬОВИХ ПАКЕТІВ} {\fbox{В.П. Лукомський},
І.С. Ганджа} {Розглянуто часову еволюцію спектрально вузьких
хвильових пакетів кінцевої амплітуди в нелінійній дисперсійній
системі, описуваній рівнянням Клейна--Ґордона з довільною
поліноміальною нелінійністю. Застосовано новий метод теорії збурень,
який дозволяє звести вихідне хвильове рівняння до модельного
рівняння для обвідної хвильового пакету (нелінійне рівняння
Шредінґера вищого порядку). Побудовано асимптотичне розвинення
часової похідної по двох незалежних параметрах, котрі характеризують
малість амплітуд ($\varepsilon$) і повільність їх змін у просторі
($\mu$). На відміну від інших методів теорії збурень (таких як метод
багатьох масштабів), де ці два параметри не розрізняють,
двопараметричний метод не приводить до появи вікових (секулярних)
доданків. Результати роботи можуть бути застосовані до дослідження
поширення ультракоротких (фемтосекундних) 
імпульсів в мережах волокнисто-оптичного зв'язку,
вивчення хвильових явищ на поверхні рідини, опису ленгмюрівських хвиль в гарячій плазмі.}

\end{multicols}
\end{document}